# Zero-shot Task Transfer for Invoice Extraction via Class-aware QA Ensemble

**Prithiviraj Damodaran**[12], **Prabhkaran Singh**[12], **Josemon Achankuju**[12]
[1] USTSmartops.ai, [2] UST Ltd

{prithiviraj.damodaran,prabhkaran.singh,josemon.achankunju}@ust.com

## Abstract

We present VESPA, an intentionally simple yet novel zero-shot system for a layout, locale and domain agnostic document extraction. In spite of the availability of large corpora of documents, the lack of labeled and validated datasets makes it a challenge to discriminatively train document extraction models for enterprises. We show that this problem can be addressed by simply *transferring* the information extraction (IE) task to a natural language Question-Answering (QA) task without engineering task-specific architectures. We demonstrate the effectiveness of our system by evaluating on a closed corpus of real-world retail and tax invoices with multiple complex layouts, domains and geographies. The empirical evaluation shows that our system outperforms 4 prominent commercial invoice solutions[1] that use discriminatively trained models with architectures specifically crafted for invoice extraction. We extracted 6 fields with zero upfront human annotation or training with an Avg. F1 of 87.50.

## 1 Introduction

Information Extraction (IE) deals broadly with the problem of extracting information from unstructured or semi-structured text. In this paper, we show that it is possible to reduce IE from documents to a problem of answering simple reading comprehension questions (Levy et al., 2017) (Figure 1) by leveraging the notion that language models are good few-shot learners (Brown et al., 2020) and the fact that a class-aware QA ensemble (Aniol et al., 2019) outperforms non-class aware ensembles and single QA models.

| Field | Sample Natural Questions |
|---|---|
| *Due date* | When is the charge due on?<br>When is the amount payable due? |
| *Invoice no* | What is the tax invoice no?<br>What is the invoice reference? |
| *Total amount* | How much is the due amount?<br>What is the total inclusive of taxes?<br>How much is the invoice amount? |

Figure 1: Fields of interest and questions with class and key phrases highlighted in blue and orange.

## 2 The research question

The underlying research question that motivates our work is: Given a set of fields of interest (FOI) like invoice amount, due date, invoice number and vendor along with configuration knobs like data types and validation policies, can we extract high quality candidates for these fields from unseen documents using a black-box ensemble of class-aware QA models in a zero-shot setting ? We choose invoice documents as they are the canonical use-case for complex form-like documents and extracting information from invoices is an actively researched area both in academia and industry (Majumder et al., 2020).

## 3 Related work

There is a long history to IE (Gaizauska et al., 1998), here are some of the most prominent techniques. Traditionally IE from documents has been framed as a template filling task (Muslea et al., 1999) or semantic slot filling task using conditional random fields (Lafferty et al., 2001) or using transformer based language models (Chen et al., 2019). However, these techniques are not directly applicable to form-like invoice documents. Early approaches in invoice extraction exploited the structure by extracting known fields based on their relative position to extracted lines (Tang et al., 1995) and detected forms (Cesarini et al., 1998). Subsequent work aims to better generalize extraction patterns by constructing formal descriptions of document structure (Coüasnon, 2006) and developing systems that allow non-expert end-users to dynamically build extraction templates ad-hoc (Schuster et al., 2013). As far as we know, in invoice extraction the following notable works were the closest to ours in terms of extraction accuracy. CloudScan system (Palm et al., 2017) framed invoice extraction as an RNN based sequence tagging task with n-gram features and representation learning. The second approach involved learning a supervised ranking model conditioned on the content of the document (Holt and Chisholm, 2018). The last approach involved learning a joint representation of the field type, neighbouring text and the potential candidates (Majumder et al., 2020) to extract high-quality candidates. But all of the above approaches used invoices in the order of 1000s for training and evaluation.

[1] Everest IDP Report, 2020

## 4 Challenges with Invoice Extraction

Building a system with wider industrial applicability that can handle all the invoice variants is a hard problem to solve because the practical challenges with invoices are at multiple levels.

**At Framing level** Invoices are complex form-like documents and the task ofextracting them sits at the intersection of natural language processing (NLP) and computer vision (CV) worlds.

**At Locale level** *Same information can be represented in different forms based on the locale or law of the land*. For instance, the concept of VAT in Europe is referred to as GST in the rest of the world. All these variants alogn with other i18n nuances like currency symbols and date formats should be factored at this level to learn good locale-specific representations.

**At Domain level** *Same information can be represented in different forms based on the business domain*. For instance, some accounts payable (AP) systems use `due_date` as an actual date, but the notion of due dates can also be net/term days from `invoice_date` in some domains. Hotel invoices have the concept of check-in / check-out date as opposed to the invoice date and due date. Field representations vary from domain to domain.

**At Data level** *To learn good general invoice representations most conventional approaches need lot of annotated and validated data.* While some of the supervised approaches can work with a relatively lower volume of human-annotated samples, most of them are really data-hungry and a critical volume of representative training samples are required to extract high quality candidates.

**At Industry level** *Adding new fields on the fly to a deployed invoice extraction solution is an industry standard requirement.* Most supervised approaches need retraining for every new field. This can be expensive and risks losing previosuly learned Human-In-The-Loop models.

**At Layout and Orientation level** *Different vendors have different invoice layouts, font styles and content positioning*. Some invoices use a columnar layout for the invoice header and borderless tables for line items, while others don't. Some documents orient text vertically.

**At Quality level** *From our corpus of tax and retail invoices, scanned invoices outnumber digital invoices by a ratio of 7:3*. Scanned documents often have quality issues like salt & pepper noise and toner issues. They can be skewed and/or can have handwritten marks like ticks or circles, watermarks and rubber stamps overlapping the invoice text. This can potentially alter the representation of the values learned by a model.

## 5 Approach overview

We look at all the above challenges through a single pane of glass and show that we can address them by creatively combining the recent Transformer based architectures for QA, semantic text retrieval, entity recognition, Object Character Recognition (OCR), image pre-processing and heuristic filtering techniques. VESPA has a document pre-processing sub-system, a document extraction subsystem, a candidate validation sub-system to eliminate wrong values or non-answers. An Expert-In-The- Loop (EITL) sub-system to handle edits and learn from it. We had to reason about this problem from the 1st principles and design a system ground-up because the conventional systems like QA4IE (Qiu et al., 2019) worked only on well-formed natural language text organized in the form of sentences and long prose like Wikipedia pages but not amenable to form-like documents like invoice.

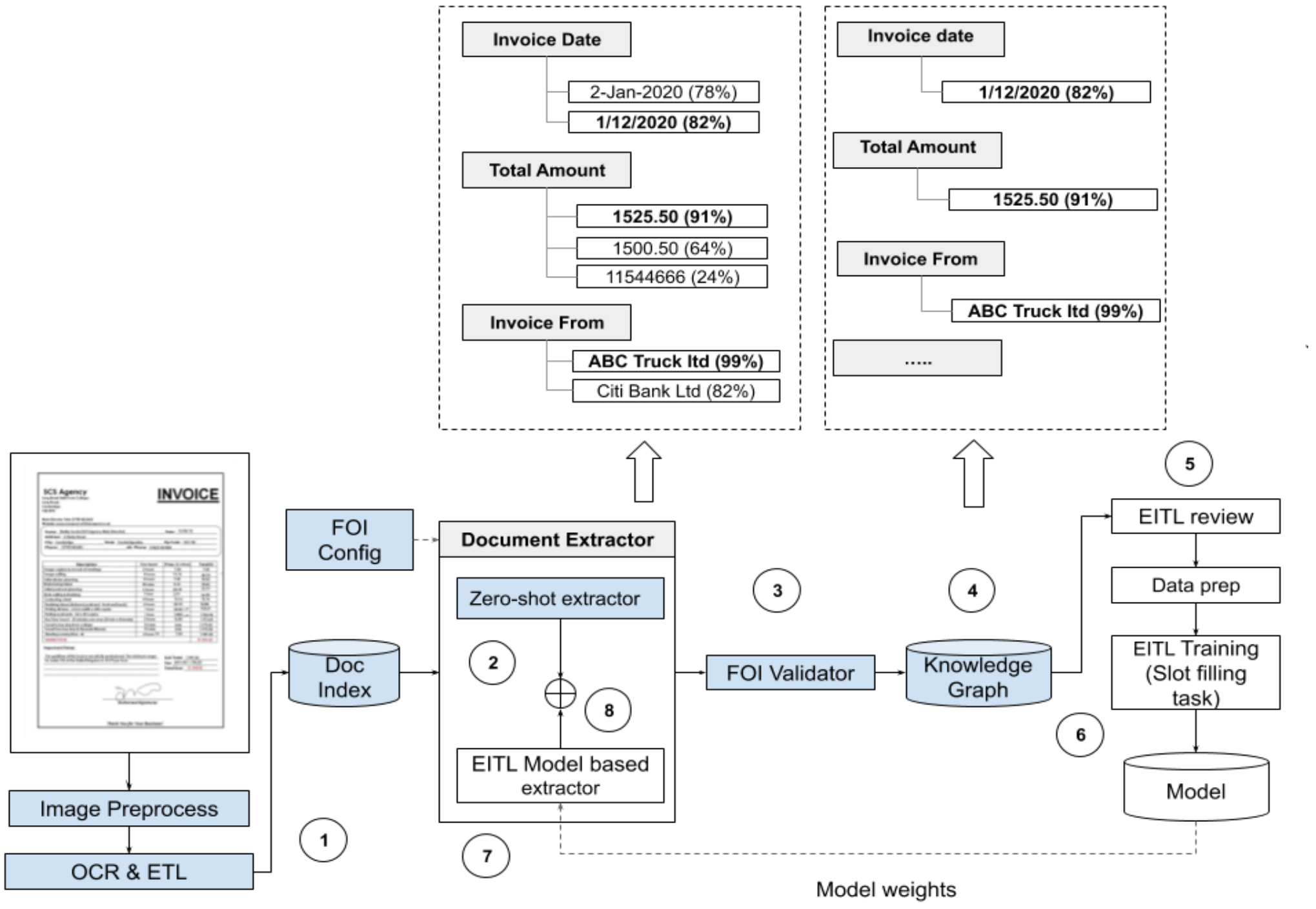

Figure 2: The end to end document extraction pipeline with the subsystems. Highlighted in blue are in the scope of this paper.

Also, taking a zero-shot approach allows us to define new fields of interest "on the fly" for a given document, without the need for retraining. This easily scales across document types, locales and domains.

## 6 VESPA Pipeline

The end to end document extraction pipeline (figure 2) is designed with a single responsibility principle (SRP) with one subsystem responsible for each step of the pipeline.

### 6.1 Image pre-processor and quality

All documents are internally converted into a high resolution, de-skewed and de-rotated TIFF image of 200 DPI to facilitate good quality OCR. For a multi-page document each page becomes a separate image. Before applying OCR the image pre-processor checks the image quality using a blind image quality assessment (IQA) technique to see if an optional pre-processing is required. We use an implementation of BRISQUE (Mittal et al., 2012) for IQA and it scores images in the range of 0 to 100, 0 being the best. Due to its distortion-genericness BRISQUE applies at pixel level for document images. Once scored we use a heuristic based approach based on the range of the scores[2] to denoise and/or sharpen the image. Post this the text-only copy of the original document is created i.e. without losing the semantics, positioning, orientation and layout of text. This is an attempt to eliminate all non-textual elements like tables, handwritten marks and logos as much as possible.

### 6.2 Extract, Transform and Load

The OCR module extracts the text from text-only images. It is pluggable, for invoice documents, we use the pre-trained Tesseract model. QA relies on efficient passage retrieval to select candidate passages for each FOI, so the extracted document contents should be indexed. We took an approach similar to haystack (Karpukhin and Oguz, 2020) and indexed the documents in ElasticSearch. But unlike haystack we index the OCR extracted contents at multiple tactical levels. Paragraph level (for Wikipedia style documents), section level (for legal documents like contracts), page level (for form-like documents e.g. bills) and table level (a CSV of line items tables) as VESPA supports a variety of document and field types. All contents are indexed as both full-text and dense vector forms. For the dense vector form we use BERT-based (Devlin et al., 2019) embeddings of 768 dimensions.

[2] Mathworks IQA scores table

| Key | Value |
|---|---|
| Locale | US *(Geo group: Country or Continent)* |
| Domain | Finance *(Domains: Legal, Healthcare)* |
| Document Type | Invoice *(Types: Receipts, Contracts)* |
| Field | Total Amount |
| Passage level | PAGE [E] (Passage for QA: can be at page, section or para levels [C]) |
| Question verbiage | { "amount due": [0.7, 0.9], "total inclusive of tax": [0.85, 0.9], "amount in dollars": [0.8, 0.8] } Learned rejection thresholds [B] |
| Question prefix | ["what is the", "How much is the"] Question = Prefix + Verbiage [A] |
| Response type | NUMERIC [D] |
| Validation policy | [ {"type": "NER", "entity": "MONEY"}, {"type": "REGEX", "pattern": "[0-9,.]+\s*USD"} ] [F] |

Table 1: Field level knobs: A sample FOI configuration

## 6.3 Field of Interest configuration

We take a declarative approach over an imperative approach (Table 1) for fields of interest (FOI). FOIs can be configured with a custom schema. Adding a new FOI is as simple as adding a new entry. VESPA supports extraction on a wide range of fields. So, for the scope of this paper and evaluation, we restrict ourselves to the following fields: Invoice_date (date), Due_date (date/days), Invoice_amount (money), Invoice_from & Invoice_to (business entity) and Invoice_number (alphanumeric string).

## 6.4 Document Extractor

The Document Extractor (DE) works like a Multi-Armed Bandit (MAB), because it uses the passage retriever to extract one or more context passages and asks all the questions for a given field to all the models in the ensemble (explore) and decides which is the best answer (exploit). Optionally, if an EITL model exists, the model inference for the given field will also be considered as possible candidate. The combiner combines all the candidate answers for the current field, validates them by passing it to the FOI validator and sorts the valid answers by the confidence score. Unlike RankQA (Kratzwald et al., 2019) we didn't have to model the passages, questions and answers to determine the top ranked answer. The candidate designated by the class-aware ensemble will be persisted as the designated value for an FOI in the knowledge graph. The EITL subsystem takes over from there.

## 6.5 Zero-shot extractor

In DE, The Zero-Shot Extractor (ZSE) asks the NL questions against the retrieved passages to an ensemble of QA models. For the FOI config shown in Table 1, ZSE will ask 6 questions (number of prefixes * number of verbiage) to each member of the class-aware ensemble. In QA, the ability to ask questions on a passage against multiple models and choosing the answers by voting for the answers based on each model's strength on that particular question-class is called class-awareness and it is a variant of weighted ensembling. We have classified the questions into 14 classes as shown in Table 2 and obtained the class specific weights for each model in our ensemble using the algorithm shown in Figure 3. The raw confidence of the extracted answers is a function of the response type of a question and its class. For instance, for numeric or money related answers a question with **"How m-m"** class yields better confidence over a question with **"what"** class. Class weightage of a model is that model's strength in answering that class of questions. For a **"what"** class question, BERT WWM **(85.81)** is a better model than BiDAF **(66.93)** as shown in Table 2. Hence, voting answers based on a weighted confidence of the answer with weight being class weight of the model helps us to extract high quality candidates.

## 6.6 Class-aware ensemble

On the choice of model architectures for the ensemble, We picked QA models with a wide range of NLU capabilities. We considered both bi-directional encoder-decoder architectures and

| Algorithm |
|---|
| **Dataset: SQuAD v2.0 evaluation set** |
| 1: **For** model *m* in the ensemble *E* **do** |
| 2: **For** class *c* in the question-classes *C* ***do*** |
| 3: $W_{cm}$ = Avg **F1** of model *m* on class *c* |

Figure 3: Get class-specific weights of models

transformers. In transformers we included both Autoencoding and Autoregressive language models to the mix. We used an optimal ensemble

WWM= Whole Word making

| Questions | | Avg. F1 of QA Models on the eval-set of SQuAD 2.0 | | | | | | |
|---|---|---|---|---|---|---|---|---|
| **Class** | **Question count** | **BERT Base** | **BiDAF** | **ELMo BiDAF** | **SpanBERT** | **BERT WWM** | **ELECTRA Large** | **BART Large** |
| Date | 22 | 77.27 | 68.18 | 72.73 | 81.82 | 72.73 | 13.64 | 90.91 |
| During | 73 | 87.67 | 73.97 | 79.45 | 84.93 | 89.04 | 6.85 | 87.67 |
| How are | 25 | 56.0 | 52.0 | 48.0 | 48.0 | 56.0 | 4.0 | 48.0 |
| How big-size | 1 | 100.0 | 0.0 | 0.0 | 0.0 | 0.0 | 0.0 | 0.0 |
| How m-m | 398 | 86.18 | 79.4 | 78.89 | 87.69 | 89.2 | 2.76 | 90.7 |
| How old | 44 | 81.82 | 72.73 | 63.64 | 88.64 | 75.0 | 2.27 | 86.36 |
| Undefined | 5 | 55.91 | 41.82 | 42.73 | 55.45 | 65.45 | 2.73 | 68.64 |
| **What** | 3910 | 78.75 | **66.93** | 70.46 | 79.34 | **85.81** | 5.96 | 86.16 |
| What time | 388 | 80.0 | 80.0 | 80.0 | 80.0 | 80.0 | 0.0 | 80.0 |
| When | 249 | 90.21 | 83.76 | 84.79 | 89.18 | 94.59 | 3.09 | 93.81 |
| Where | 12 | 78.31 | 68.27 | 69.08 | 75.9 | 85.14 | 4.02 | 84.34 |
| Who | 487 | 85.01 | 80.29 | 82.14 | 82.14 | 88.3 | 6.78 | 91.17 |
| Whom | 94 | 91.67 | 75.0 | 83.33 | 75.0 | 83.33 | 16.67 | 91.67 |
| Why | 220 | 48.94 | 46.81 | 40.43 | 59.57 | 58.51 | 2.13 | 61.7 |

Table 2: Class-awareness: Class weights for Models in Ensemble QA

search technique similar to the one shown by Jiwei et al., 2020 to arrive the list of models BIDAF (Seo et al., 2016), ELMo (Peters et al., 2018), SpanBERT (Joshi et al., 2019), ELECTRA (Clark et al., 2020), BART (Lewis et al., 2019).

### 6.7 Field of Interest Validator

A Class-aware QA ensemble is very good in getting high quality candidates for fields if the answer exists. But QA models are extractive by design and so cannot say "answer is not found" i.e. the field cannot be extracted. This gets tricky with very similar fields like invoice date and due date. Most QA models will bring invoice date or any other date as an answer for a due date question even if due date doesn't exist in the document. So, rejecting non-answers answers is crucial to VESPA and it happens at multiple levels. Firstly we validate and reject all answers based on the data type configured. Secondly, we reject answers using the thresholds we learned based on the answer response type, question class and model weight for each class. For instance, even if one or more QA models bring the invoice date instead of due date in the absence of due date it will get filtered out at the threshold level. Because all the QA models were fine-tuned on SQuAD 2.0 (Rajpurkar et al., 2018) and it has adversarial questions crafted to help QA models learn to abstain from answering questions wherever not possible. Hence wrong answers usually come with lower confidence. On the off-chance when non-answers do pass through the confidence threshold filter, that's where the FOI validator comes in. It applies the list of validation policies configured for a field (as shown Table 1F) and filters answers which don't pass the validations. We use different types of validation policies to validate a range of fields. For instance, the named-entity recognition (NER) policy is used for standard fields like ORG, DATE and MONEY. We use both pretrained and custom NER models. Specific string patterns like SSN or Phone numbers can be validated using a REGEX policies. Upon validation the confidence score is boosted by a configurable factor.

### 6.8 Pre and Post processors

VESPA supports Snorkel style (Ratner et al., 2017) pluggable pre and post-processors. Validated answers can be optionally post-processed with custom logic as drop-in features. Document extractor calls custom processors before and after the FOI extraction. For instance, custom post-processors for invoice transforms all dates to yyyy-mm-dd (ISO date format).

## 7 Empirical evaluation and results

We evaluated both the single QA (BERT base) and the class-ware ensemble QA versions (models in listed Table 2) of VESPA alongside few prominent commercial solutions as shown by Holt and Chisholm, 2018 and Everest Intelligent document processig 2020 report to prove that in extracting unstructured information from documents: A zero-shot task transfer of information extraction (IE) to a class-aware ensemble QA can match and outperform heavily-engineered task specific architectures (Table 3).

EB= Ezzy Bills, AFR = Azure Form Recognizer, CAE = Class-Aware Ensemble, NS=Not Supported

| Fields | EB | SYPHT | AFR | Rossum | Ours | |
|---|---|---|---|---|---|---|
| | | | | | Single QA | CAE QA |
| Invoice Date | 66 | 66 | 77 | 79 | 72 | **81** |
| Invoice From | 24 | *NS* | 83 | 74 | 87 | **90** |
| Due Date | 26 | 45 | 55 | 83 | 82 | **88** |
| Invoice Amount | 91 | 85 | 79 | **94** | 84 | 90 |
| Invoice Number | 81 | 83 | 66 | **93** | 81 | 79 |
| Invoice To | *NS* | *NS* | 78 | 88 | 89 | **97** |
| **Avg. F1** | 58.0 | 70.0 | 73.0 | 85.0 | 82.67 | **87.50** |

Table 3: Extraction accuracy by field of interest on Invoice Documents

We chose EzzyBills[3], SYPHT[4], Rossum[5] and Azure Form Recogniser[6] and all of them offer open APIs to train (optionally) and extract fields. We could not consider solutions like ABBYY FlexiCapture due to lack of open APIs. We evaluated all the solutions on **300** documents from a private corpus of retail and tax invoice documents with multiple complex layouts from domains like logistics, telecommunication and manufacturing from across geographies. We reused the standard F1 metric used for the QA tasks for this comparison. The empirical evidence shows that our system outperforms solutions with discriminatively trained models specifically crafted for invoice extraction on multiple fields. Both single QA version and the class-aware ensemble QA version of our system rankly outperforms other solutions on the tricky due-date field. Due to the intellectually property nature of our solution we can neither publicly share the VESPA code nor the private datasets.

## 8 Future work

The focus of the future work has two main themes: Lifting extraction accuracy and Reducing model inference time for extraction. As for as extraction accuracy goes we would like to try the following 1.) Use pretrained QA models in a class-aware mixture-of-experts setting (Jacobs et al., 1998) as opposed to a weighted ensemble. 2.) Fine tune the QA models with BoolQ (Clark et al., 2019) dataset to device Yes/No questions to know if a field exists in a document or not 3.) Include QA models trained on other datasets like DROP (Dua et al., 2019) and RACE (Lai and Xie, 2017) 4.) Automatically generate questions with document text as context as the quality of the fields extracted largely depends on the quality of the questions. 5.) Strengthen the domain awareness by using domain-specific pretrained models like FinBERT (Yang et al., 2020), ClinicalBERT (Huang et al., 2019) for respective document types. In the inference time front we would like to 6.) Port pretrained models using ONNX and 7.) Try some model quantisation, pruning and distillation to reduce the overall model size and improve inference speed.

## 9 Conclusion

Our experiments confirm that a class-aware QA ensemble based on a good mix of auto-encoding and auto-regressive pretrained language models can outperform heavily-engineered task specific architectures in extracting high quality information from documents in a zero-shot setting. The empirical evaluation shows that our intentionally simple framework rankly outperforms 4 commercial invoice solutions on 4 out of the 6 key fields. **VESPA**: **V**ery **E**mbarrassingly **S**imple yet **P**henomenally **A**ccurate.

[3] https://www.ezzybills.com/
[4] https://www.sypht.com/
[5] https://rossum.ai/,
[6] Azure form recogniser